\documentclass[12pt]{article}
\usepackage{amssymb}
\usepackage{amsfonts}

\input{tcilatex}
\begin{document}

\begin{center}
\bigskip

\textbf{The Lorentz group and its finite field analogues: }

\textbf{local isomorphism and approximation}
\end{center}

\bigskip

\begin{center}
Stephan Foldes

Tampere University of Technology

PL 553, 33101 Tampere, Finland

sf@tut.fi

4 August 2008

\bigskip

\textbf{Abstract}
\end{center}

\textit{Finite Lorentz groups acting on 4-dimensional vector spaces
coordinatized by finite fields with a prime number of elements are
represented as homomorphic images of countable, rational subgroups of the
Lorentz group acting on real 4-dimensional space-time. Bounded subsets of
the real Lorentz group are retractable with arbitrary precision to finite
subsets of such rational subgroups. These finite retracts correspond, via
local isomorphisms, to well-behaved subsets of Lorentz groups over finite
fields. This establishes a relationship of approximation between the real
Lorentz group and Lorentz groups over very large finite fields.}

\textit{\bigskip }

\bigskip

\textbf{1 Finite Minkowski spaces and statement of the theorem\bigskip }

\bigskip

The purpose of this paper is to establish a relationship of approximation
between Lorentz transformations of Minkowski space $%
\mathbb{R}
^{4}$ and Lorentz transformations of 4-dimensional spaces over very large
finite fields, in the precise sense of Theorem 1 below. In relation to
theoretical alternatives in particle physics, Lorentz transformations over
finite fields seem to have been first considered by Coish in [C], then by
Shapiro [S], Ahmavaara [A1, A2], Yahia [Y], Joos [Jo], Beltrametti and Blasi
[BB1], and Nambu [N]. Much of the algebra of Lorentz groups over finite
fields had already been developed by Dickson [D], and in the context of
physical applications this theory was brought up to date by Beltrametti and
Blasi [BB2]. A finite analogue of the Alexandrov-Zeeman characterization of
the Lorentz group via optical causality was established by Blasi, Gallone,
Zecca and Gorini [BGZG]. The approximation described in Theorem 1 relies on
the one hand on modifying a relationship of approximation between positive
real numbers and quadratic residues modulo a large prime established by
Kustaanheimo [JK, K1, K2], and modifying J\"{a}rnefelt's approximation of
collinear ranges of points in Euclidean space by collinear ranges in finite
geometries [J\"{a}]. The proof of Theorem 1 also makes essential use of
rotation-boost decomposition, and of representing Lorentz groups over finite
fields as quotients of appropriate subgroups of the real Lorentz group. The
algebraic background and terminology used in this theorem are as follows.

\bigskip

We consider fields $\mathbf{F}$ in which $-1$ is not a square, and we shall
assume throughout that $\mathbf{F}$ has this property. (The main cases of
interest are the field $%
\mathbb{R}
$ of real numbers and the $p$-element finite fields $\mathbf{F}_{p}$ for
prime numbers $p\equiv 7\func{mod}8$, although many of the basic facts are
also true for $p$-element fields with primes $p\equiv 3\func{mod}8$.) For
any such field $\mathbf{F}$ the $4$-dimensional vector space $\mathbf{F}^{4}$
is endowed with the ($\mathbf{F}$-\textit{valued}) \textit{Minkowski norm} 
\[
\mu (t,x,y,z)=t^{2}-x^{2}-y^{2}-z^{2} 
\]%
A \textit{proper} \textit{Lorentz transformation over} $\mathbf{F}$ is a
linear transformation $T$ from the vector space $\mathbf{F}^{4}$ to itself
which preserves the Minkowski norm (\textit{i.e.,} $\mu (T(\mathbf{v))}=\mu (%
\mathbf{v)}$ for all $\mathbf{v}\in \mathbf{F}^{4}$) and which has
determinant $1$. For any given field $\mathbf{F}$, these transformations
form a group that we call the \textit{proper Lorentz group over} $\mathbf{F}$%
, denoted $\mathcal{L}_{+}\mathbf{F}$. If the field $\mathbf{F}$ is not
specified, it is understood to be $%
\mathbb{R}
.$ A Lorentz transformation over a field $\mathbf{F}$ is said to be \textit{%
orthochronous} if it maps $(1,0,0,0)$ to a vector $(t,x,y,z)$ where $t$ is a
non-zero square in $\mathbf{F}$. For $\mathbf{F}=%
\mathbb{R}
$ orthochronous Lorentz transformations form a subgroup of $\mathcal{L}_{+}%
\mathbb{R}
$ called the \textit{orthochronous proper Lorentz group, }denoted\textit{\ }$%
\mathcal{L}_{+}^{\uparrow }%
\mathbb{R}
.$ Over finite fields the composition of orthochronous transformations is
not necessarily orthochronous.

\bigskip

If $G$ and $H$ are two (abstract) groups and $A\subseteq G,$ $Y\subseteq H$
are arbitrary sets of group elements, a \textit{local isomorphism} between $%
A $ and $Y$ is a bijective map $\sigma $ from $A\cup AA$ to $Y\cup YY$
(where $AA=\left\{ xy:x,y\in A\right\} $) such that $\sigma \lbrack A]=Y$
and for all $x,y\in A$%
\[
\sigma (xy)=\sigma (x)\sigma (y) 
\]%
It follows that if $1_{G}\in A$ then $\sigma (1_{G})=1_{H\text{ \ }}$and if $%
x,x^{-1}\in A$ then $\sigma (x^{-1})=\sigma (x)^{-1}.$ If such a local
isomorphism exists, the sets $A$ and $Y$ are said to be \textit{locally
isomorphic. }(If $A$ and $Y$ are subgroups they are locally isomorphic if
and only if they are isomorphic, but otherwise the sets $A$ and $Y$ may be
locally isomorphic even if the subgroups they generate are not isomorphic.)

\bigskip

By the \textit{norm} $\left\Vert T\right\Vert $\ of any linear
transformation $T:%
\mathbb{R}
^{4}\longrightarrow 
\mathbb{R}
^{4}$ with standard matrix representation $(a_{ij})_{1\leq i,j\leq 4}$ we
mean $(\Sigma a_{i,j}^{2})^{1/2}$, i.e. the Euclidean ($l_{2}$) norm of the
matrix. By a \textit{retraction of} $\mathcal{L}_{+}^{\uparrow }\mathbf{%
\mathbb{R}
}$ \textit{to a subset} $A\subseteq \mathcal{L}_{+}^{\uparrow }\mathbf{%
\mathbb{R}
}$ we mean a map $f:\mathcal{L}_{+}^{\uparrow }\mathbf{%
\mathbb{R}
}\longrightarrow A$ such that $f(T)=T$ for all $T\in A$ (\textit{i.e.,} such
that $f^{2}=f$).

\bigskip

\textbf{Theorem 1 }\ \textit{For every} $\epsilon >0$ \textit{and }$M>0$ 
\textit{the orthochronous proper Lorentz group} $\mathcal{L}_{+}^{\uparrow }%
\mathbb{R}
$ \textit{has a retraction }$\ f:\mathcal{L}_{+}^{\uparrow }%
\mathbb{R}
\longrightarrow A$ \textit{to a finite subset} $A\subseteq \mathcal{L}%
_{+}^{\uparrow }%
\mathbb{R}
$ \textit{satisfying}

\[
\left\Vert T-f(T)\right\Vert <\epsilon 
\]%
\textit{for all Lorentz transformations} $T\in \mathcal{L}_{+}^{\uparrow }%
\mathbb{R}
$ \textit{of norm not exceeding} $M$, \textit{and such that} \textit{for
some prime number} $p\equiv 7\func{mod}8$ \textit{the subset} $A$\textit{\
is locally isomorphic to a set }$Y$ \textit{of orthochronous transformations
in the proper Lorentz group }$\mathcal{L}_{+}\mathbf{F}_{p}$\textit{\ over
the finite field }$\mathbf{F}_{p}.$

\bigskip

The proof is provided in Section 2, together with some elaborations of the
mathematical aspects that are needed for the sake of precision and clarity.
These are presented as a succession of intermediate propositions. A
generalization to the extended Lorentz group is presented in Theorem 2 of
Section 3, with some additional properties of the local isomorphism linking
the respective Lorentz groups over the real numbers and over finite fields.
In Section 4 we briefly comment on some of the particularities of finite
analogues of the Lorentz group.

\bigskip

\bigskip

\textbf{2 Background developments and proof of Theorem \bigskip 1}\bigskip

\textbf{2.1 Proper Lorentz groups over fields where -1 is not a square}

\bigskip

Much of the basic theory of finite Lorentz groups that we need is contained
in or follows from the seminal work of Dickson [D] and the more recent study
of Beltrametti and Blasi [BB2].

\bigskip\ 

Lorentz groups are defined according to the active perspective as groups of
bijective transformations, and composition of transformations is denoted
simply by juxtaposition $T_{1}T_{2}$, where $T_{2}$ is the transformation
first applied. If the standard matrix representations of these
transformations are $\mathbf{T}_{1}$ and $\mathbf{T}_{2}$ then the matrix
product $\mathbf{T}_{1}\mathbf{T}_{2}$ represents the transformation $%
T_{1}T_{2}.$

A proper Lorentz transformation $T$ over a field $\mathbf{F}$ where $-1$ is
not a square is called a \textit{space rotation} (\textit{over} $\mathbf{F}$%
) if it fixes the first standard basis vector $(1,0,0,0)\in $ $\mathbf{F}%
^{4} $. Equivalently, space rotations over $\mathbf{F}$ are the linear
transformations of $\mathbf{F}^{4}$ with standard matrix representation of
the block diagonal form%
\[
\left( 
\begin{array}{cc}
1 &  \\ 
& \mathbf{R}_{3}%
\end{array}%
\right) 
\]%
where $\mathbf{R}_{3}$ is a 3-by-3 orthogonal matrix (i.e. $\mathbf{R}%
_{3}^{\top }=\mathbf{R}_{3}^{-1}$) with determinant $1$. There are three
space rotations which permute the standard basis vectors, they are given by%
\[
\mathbf{R}_{3}=\left( 
\begin{array}{ccc}
1 &  &  \\ 
& 1 &  \\ 
&  & 1%
\end{array}%
\right) ,\mathbf{R}_{3}=\left( 
\begin{array}{ccc}
& 1 &  \\ 
&  & 1 \\ 
1 &  & 
\end{array}%
\right) ,\mathbf{R}_{3}=\left( 
\begin{array}{ccc}
&  & 1 \\ 
1 &  &  \\ 
& 1 & 
\end{array}%
\right) 
\]%
These particular space rotations are called \textit{rotations of standard
space axes} and they form a $3$-element subgroup of $\mathcal{L}_{+}\mathbf{F%
}$. All space rotations constitute a subgroup of $\mathcal{L}_{+}\mathbf{F}$
over any field $\mathbf{F.}$ If a space rotation fixes one of the three 
\textit{standard space unit vectors} $(0,1,0,0),(0,0,1,0)$ or $(0,0,0,1)$
then it is said to be an \textit{elementary space rotation} (\textit{around}
that unit vector). For each of these three unit vectors, elementary space
rotations around that vector constitute a subgroup of the group of all space
rotations. Around $(0,1,0,0)$ this subgroup consists of the identity, the 
\textit{half-turn around} $(0,1,0,0)$ given by%
\[
(t,x,y,z)\mapsto (t,x,-y-z) 
\]%
and for each $\alpha \neq 0$ in $\mathbf{F}$ the transformation $R_{\alpha }$
called a \textit{basic rotation} whose standard matrix representation is 
\[
\left( 
\begin{array}{cccc}
1 &  &  &  \\ 
& 1 &  &  \\ 
&  & \frac{\alpha -\alpha ^{-1}}{\alpha +\alpha ^{-1}} & \frac{2}{\alpha
+\alpha ^{-1}} \\ 
&  & \frac{-2}{\alpha +\alpha ^{-1}} & \frac{\alpha -\alpha ^{-1}}{\alpha
+\alpha ^{-1}}%
\end{array}%
\right) 
\]%
Each of the two groups of elementary space rotations around $(0,0,1,0)$ and $%
(0,0,0,1)$ is obtained from the group of elementary space rotations around $%
(0,1,0,0)$ by conjugation with space axis rotations represented by
appropriate permutation matrices. All space rotations are orthochronous,
over any field $\mathbf{F.}$

For every non-zero square $\alpha $ in $\mathbf{F}$ (which means positive $%
\alpha $ in $%
\mathbb{R}
$, quadratic residue class in $%
\mathbb{Z}
/p%
\mathbb{Z}
\cong \mathbf{F}_{p}$) the \textit{basic boost} $B_{\alpha }$ is defined as
the Lorentz transformation over $\mathbf{F}$ with standard matrix
representation 
\[
\left( 
\begin{array}{cccc}
\frac{\alpha +\alpha ^{-1}}{2} & \frac{\alpha -\alpha ^{-1}}{2} &  &  \\ 
\frac{\alpha -\alpha ^{-1}}{2} & \frac{\alpha +\alpha ^{-1}}{2} &  &  \\ 
&  & 1 &  \\ 
&  &  & 1%
\end{array}%
\right) 
\]%
For any space rotation $R$ and $\alpha $ any non-zero square in $\mathbf{F}$%
, the conjugate $RB_{\alpha }R^{-1}$ is called a \textit{boost, }and for any
given space rotation $R$\ the set of boosts 
\[
\{RB_{\alpha }R^{-1}:\alpha \neq 0,\text{ }\alpha \text{ square in }\mathbf{F%
}\} 
\]%
is a subgroup of $\mathcal{L}_{+}\mathbf{F}$. If the space rotation $R$ is a
rotation of standard space axes, then $RB_{\alpha }R^{-1}$ is said to be an 
\textit{elementary boost}. A boost is elementary if and only if it fixes two
of the three basis vectors $(0,1,0,0),(0,0,1,0),(0,0,0,1)$. All boosts are
orthochronous over $%
\mathbb{R}
,$\textbf{\ }but boosts over a finite field do not need to be orthochronous:
consider over $\mathbf{F}_{7}$ any of the two non-trivial boosts $B_{2}$ and 
$B_{4}.$

\bigskip

Over any field $\mathbf{F}$, note that composing \textit{space-time reversal}
\[
(t,x,y,z)\mapsto (-t,-x,-y,-z) 
\]%
with the half-turn around $(0,1,0,0)$ yields \textit{reflection in the yz
plane }given by%
\[
(t,x,y,z)\mapsto (-t,-x,y,z). 
\]%
Basic boosts together with this reflection generate the subgroup of $%
\mathcal{L}_{+}\mathbf{F}$ fixing $(0,0,1,0)$ and $(0,0,0,1)$. (This
subgroup is also referred to as the group of hyperbolic isometries of the 
\textit{tx} plane.)

\bigskip

Since every space rotation in $\mathcal{L}_{+}\mathbf{%
\mathbb{R}
}$ is the product of $3$ elementary space rotations, elementary space
rotations and basic boosts generate $\mathcal{L}_{+}^{\uparrow }\mathbf{%
\mathbb{R}
}$, and together with reflection in the \textit{yz} plane they generate $%
\mathcal{L}_{+}\mathbf{%
\mathbb{R}
.}$ On the other hand, a classical result on linear groups over finite
fields due to Dickson [D] states, when particularized to the case of
4-dimensional spaces over $\mathbf{F}_{p}$ for primes $p\equiv 7\func{mod}8$%
, that $\mathcal{L}_{+}\mathbf{F}_{p}$ is generated by elementary rotations,
elementary boosts and reflections in the \textit{yz} plane. Re-stated in a
slightly strengthened form, based on the observation that all elementary
boosts are conjugates of basic boosts by axis rotations which are in turn
products of elementary space rotations, we have the following:

\bigskip

\textbf{Rotation-Boost Lemma over Finite Fields }(from Dickson [D]) \textit{%
For any prime number }$p\equiv 7\func{mod}8,$ \textit{the proper Lorentz
group} $\mathcal{L}_{+}\mathbf{F}_{p}$ \textit{is generated by basic boosts,
elementary space rotations and space-time reversal.\ \ }$\square $

\bigskip

In fact the above lemma also holds for primes $p\equiv 3\func{mod}8$ with
the exception of $p=3$ (see [D]) but this fact is not needed in what follows.

\bigskip

Unlike over the real number field $%
\mathbb{R}
$, over finite fields not every proper Lorentz transformation can be
represented in the form $RB$ (or $BR)$ with a rotation $R$ and a boost $B.$
(Any such product of a rotation and a boost transforms $(1,0,0,0)$ to a
vector $(t,x,y,z)$ where $x^{2}+y^{2}+z^{2}$ is a square in the field of
scalars. However, over finite fields not all proper Lorentz transformations,
not even the orthochronous ones, have this property.) For the real case see
Moretti's study [M]\ for a comprehensive perspective on rotation-boost
decomposition.

\bigskip

\textbf{2.2 Lorentz transformations over local rings}

\bigskip

We represent the $p$-element field $\mathbf{F}_{p}$ as the quotient of the
localization ring 
\[
\mathcal{%
\mathbb{Z}
}_{(p)}=%
\mathbb{Z}
\cdot (%
\mathbb{Z}
\diagdown p%
\mathbb{Z}
)=\{st^{-1}:s,t\in 
\mathbb{Z}
,\text{ \ }t\neq 0,\text{ }(p,t)=1\} 
\]%
by its unique maximal ideal $p\mathcal{%
\mathbb{Z}
}_{(p)}$, an approach also taken by Ahmavaara [A1].\ There is a unique
surjective ring homomorphism from the localization ring $\mathcal{%
\mathbb{Z}
}_{(p)}$ onto the $p$-element field $\mathbf{F}_{p}$, called the \textit{%
canonical map} (or \textit{canonical surjection}) from $%
\mathbb{Z}
_{(p)}$ to $\mathbf{F}_{p}$. (The restriction of the canonical map to $%
\mathbb{Z}
\subseteq $ $\mathcal{%
\mathbb{Z}
}_{(p)}$ is the unique surjective ring homomorphism $%
\mathbb{Z}
\longrightarrow \mathbf{F}_{p}$ corresponding to the usual representation of 
$\mathbf{F}_{p}$ as a quotient of $%
\mathbb{Z}
.$) The canonical map from $%
\mathbb{Z}
_{(p)}$ to $\mathbf{F}_{p}$\ \textit{induces} a surjective ring homomorphism
from the ring of $4$-by-$4$ matrices with entries in $\mathcal{%
\mathbb{Z}
}_{(p)}$ onto the ring of $4$-by-$4$ matrices with entries in $\mathbf{F}%
_{p} $ and thus \textit{induces} a map from any set of linear
transformations of $%
\mathbb{R}
^{4}$\ with coefficients in $\mathcal{%
\mathbb{Z}
}_{(p)}$ to the set of linear transformations of$\ \mathbf{F}_{p}^{4}$ \ 

\bigskip

For any integral domain $\mathbf{D}$ in which $-1$ is not a square, those $%
\mathbf{D}$-module automorphisms $L$ of $\mathbf{D}^{4}$ whose standard $%
4\times 4$ matrix representation has determinant $1$ and which leave
invariant the ($\mathbf{D}$-\textit{valued) Minkowski norm}

\[
\mu (t,x,y,z)=t^{2}-x^{2}-y^{2}-z^{2} 
\]%
are called \textit{Lorentz transformations} \textit{over} $\mathbf{D}$ and
they constitute a group, called the \textit{proper} \textit{Lorentz group
over} $\mathbf{D}$, denoted $\mathcal{L}_{+}\mathbf{D.}$ A transformation $L$
in $\mathcal{L}_{+}\mathbf{D}$ is said to be \textit{orthochronous} if it
maps $(1,0,0,0)$ to a vector $(t,x,y,z)$ such that $t$ is a non-zero square
in $\mathbf{D}$. The set $\mathcal{L}_{+}^{\uparrow }\mathbf{D}$ of such
orthochronous transformations may or may not be a subgroup of $\mathcal{L}%
_{+}\mathbf{D}$. A $\mathbf{D}$-module automorphism $L$ of \ $\mathbf{D}^{4}$
having determinant $1$ belongs to $\mathcal{L}_{+}\mathbf{D}$ if and only if
its standard matrix representation $\mathbf{L}$ and the diagonal matrix $%
\mathbf{J}$ with main diagonal $(1,-1,-1,-1)$ satisfy the equation $\mathbf{%
LJL}^{\top }=\mathbf{J.}$ For any set $C$ of elements of the integral domain 
$\mathbf{D,}$ we denote by $\mathcal{L}_{+}\mathbf{D}C$, respectively by $%
\mathcal{L}_{+}^{\uparrow }\mathbf{D}C,$ the set of those $L\in \mathcal{L}%
_{+}\mathbf{D,}$ respectively $L\in \mathcal{L}_{+}^{\uparrow }\mathbf{D,}$
for which all entries of the standard matrix representation of $L$ are in\ $%
C $. If $\mathbf{C}$ is a subring of $\mathbf{D}$ (containing the unit $1$
of $\mathbf{D}$) then $\mathcal{L}_{+}\mathbf{DC}$ is a subgroup of $%
\mathcal{L}_{+}\mathbf{D}$. and restricting transformations $L\in \mathcal{L}%
_{+}\mathbf{DC}$ to $\mathbf{C}^{4}\subseteq \mathbf{D}^{4}$ yields a group
isomorphism, called\textit{\ canonical isomorphism,}%
\[
\rho :\mathcal{L}_{+}\mathbf{DC}\longrightarrow \mathcal{L}_{+}\mathbf{C} 
\]%
(The closure of $\mathcal{L}_{+}\mathbf{DC}$ under inversion is due to the
fact that for all $L\in \mathcal{L}_{+}\mathbf{D}$ with matrix $\mathbf{L}$
we have $\mathbf{L}^{-1}\mathbf{=JL}^{\top }\mathbf{J}$.) The isomorphism $%
\rho :\mathcal{L}_{+}\mathbf{DC}\longrightarrow \mathcal{L}_{+}\mathbf{C}$
obviously maps orthochronous transformations to orthochronous
transformations, $\rho \lbrack \mathcal{L}_{+}^{\uparrow }\mathbf{DC}]=%
\mathcal{L}_{+}^{\uparrow }\mathbf{C,}$ and\ $\mathcal{L}_{+}^{\uparrow }%
\mathbf{DC}$ and $\mathcal{L}_{+}^{\uparrow }\mathbf{C}$ are locally
isomorphic subsets even if they are not subgroups of $\mathcal{L}_{+}\mathbf{%
DC}$ and $\mathcal{L}_{+}\mathbf{C.}$ They are isomorphic subgroups,
however, in the case of particular interest, which is $\mathbf{D}=%
\mathbb{R}
$ and $\mathbf{C}=\mathcal{%
\mathbb{Z}
}_{(p)}$ (for any $p\equiv 3\func{mod}4).$ The following lemma provides a
representation of the proper Lorentz group over a finite field $\mathbf{F}%
_{p}$ as a quotient of the proper Lorentz group over the localization ring $%
\mathbb{Z}
_{(p)},$ or equivalently, as a quotient of the group of (real) proper
Lorentz transformations with coefficients in $%
\mathbb{Z}
_{(p)}.$

\bigskip

\textbf{Homomorphism Lemma }\textit{Let }$p$\textit{\ be any prime number
congruent to }$7$ \textit{modulo} $8.$ \textit{The canonical map }$%
\mathbb{Z}
_{(p)}$\textit{\ }$\longrightarrow \mathbf{F}_{p}$\textit{\ induces a
surjective group homomorphism }$\mathcal{L}_{+}%
\mathbb{Z}
_{(p)}\longrightarrow \mathcal{L}_{+}\mathbf{F}_{p}$\textit{. Pre-composing
it with the canonical isomorphism }$\rho $ \textit{yields a surjective group
homomorphism }$\mathcal{L}_{+}%
\mathbb{R}
\mathcal{%
\mathbb{Z}
}_{(p)}\longrightarrow \mathcal{L}_{+}\mathbf{F}_{p}.$

\textbf{Proof\ \ }Its is obvious that the map $\mathcal{L}_{+}%
\mathbb{Z}
_{(p)}\longrightarrow \mathcal{L}_{+}\mathbf{F}_{p}$ induced by the
canonical map $%
\mathbb{Z}
_{(p)}$\textit{\ }$\longrightarrow \mathbf{F}_{p}$ is a group homomorphism.
In $\mathcal{L}_{+}\mathbf{F}_{p}$ every basic boost, every elementary
rotation and also the space-time reversal transformation is in the range of
this homomorphism. Its surjectivity is then a consequence of the
Rotation-Boost Lemma over Finite Fields. $\ \ \ \ \ \ \ \ \ \ \ \ \ \ \ \ \
\ \ \ \ \ \ \ \ \ \square $

\bigskip

The two homomorphisms onto $\mathcal{L}_{+}\mathbf{F}_{p}$ defined by the
above lemma will be also referred to as \textit{canonical}.

\bigskip

\bigskip

\textbf{2.3 Local isomorphisms}\bigskip

The next few lemmas are easily verified.

\bigskip

\textbf{Injection Lemma 1 \ }\textit{If }$h:G_{1}\longrightarrow G_{2}$ 
\textit{is a group homomorphism,} $A\subseteq G_{1}$\textit{\ and the
restriction} $h_{A}$ \textit{of} $h$ \textit{to} $A\cup AA$ \textit{is} 
\textit{injective, then} $h_{A}$ \textit{is a local isomorphism between} $A$ 
\textit{and }$h\left[ A\right] .$ \ \ \ \ \ \ \ \ \ \ \ \ \ \ \ \ \ \ \ \ \
\ \ \ \ \ \ \ \ \ \ \ \ \ \ \ \ \ \ \ \ \ \ \ \ \ \ \ \ \ \ \ \ \ \ \ \ \ \
\ \ \ \ \ \ \ \ \ \ \ \ \ \ \ \ \ \ \ \ \ \ \ \ \ $\square $

\bigskip

Ahmavaara [A1] seems to have been aware of the fact stated in Injection
Lemma 1, at least in the particular case where $h$ is the canonical
surjection $%
\mathbb{Z}
_{(p)}\longrightarrow \mathbf{F}_{p}$ viewed as an additive group
homomorphism, while apparently disregarding (or treating differently) the
case where $h$ is the restriction of the same canonical map to $%
\mathbb{Z}
_{(p)}^{\ast }=%
\mathbb{Z}
_{(p)}\setminus p%
\mathbb{Z}
_{(p)}$ and viewed as a multiplicative group homomorphism onto $\mathbf{F}%
_{p}^{\ast }=\mathbf{F}_{p}\setminus \{0\}.$

For each positive integer $k$ consider the set $C_{k}$ of rational numbers
defined by 
\[
C_{k}=\{st^{-1}:s,t\in 
\mathbb{Z}
,\text{ \ }t\neq 0,\left\vert s\right\vert \leq k,\text{ }\left\vert
t\right\vert \leq k\} 
\]%
If $p$ is a prime number larger than $k,$ then $C_{k}\subseteq \mathcal{%
\mathbb{Z}
}_{(p)}.$

\textbf{\ }

\textbf{Injection Lemma 2}\textit{\ \ For any prime number }$p\equiv 3\func{%
mod}4$\textit{\ let} $k$ \textit{be a positive integer such that} $2k^{2}<p$%
. \textit{Then the canonical map }$%
\mathbb{Z}
_{(p)}\longrightarrow \mathbf{F}_{p}$ \textit{is injective on }$C_{k}$ 
\textit{and the canonical map} $\mathcal{L}_{+}%
\mathbb{R}
\mathbb{Z}
_{(p)}\longrightarrow \mathcal{L}_{+}\mathbf{F}_{p}$ \textit{is injective on}
$\mathcal{L}_{+}%
\mathbb{R}
C_{k}.$ \ \ \ \ \ \ \ \ \ $\square $

\bigskip

\textbf{Injection Lemma 3}\textit{\ \ For any prime number }$p\equiv 3\func{%
mod}4$ \textit{let} $k$ \textit{be a positive integer such that} $32k^{16}<p$%
. \textit{Let }$A=\mathcal{L}_{+}%
\mathbb{R}
C_{k}$. \textit{Then the canonical homomorphism }$\mathcal{L}_{+}%
\mathbb{R}
\mathbb{Z}
_{(p)}\longrightarrow \mathcal{L}_{+}\mathbf{F}_{p}$ \textit{is injective on 
}$A\cup AA.$

\textbf{Proof} \ $A\cup AA\subseteq \mathcal{L}_{+}%
\mathbb{R}
C_{4k^{8}}$ and Injection Lemma 2. \ \ \ $\square $

\bigskip

These yield the following

\bigskip

\textbf{Local Isomorphism Lemma}\textit{\ \ For every positive integer k and
prime number p}$\equiv 3\func{mod}4$ \textit{\ such that\ }$p>32k^{16}$, 
\textit{the set} $\mathcal{L}_{+}%
\mathbb{R}
C_{k}\mathit{\ }$\textit{of proper Lorentz transformations with coefficients
in }$C_{k}$ \textit{is locally isomorphic to a subset of the finite proper
Lorentz group }$\mathcal{L}_{+}\mathbf{F}_{p}$. \textit{A local isomorphism
is provided by restricting the canonical homomorphism} $\mathcal{L}_{+}%
\mathbb{R}
\mathbb{Z}
_{(p)}\longrightarrow \mathcal{L}_{+}\mathbf{F}_{p}$ \textit{to} $\mathcal{L}%
_{+}%
\mathbb{R}
C_{k}.$ $\ \ \ \square $

\bigskip

Using Dirichlet's theorem on primes in arithmetic progressions, in [K1]
Kustaanheimo proved the following:

\bigskip

\textbf{Kustaanheimo's Chain Theorem }[K1] \textit{For every positive
integer }$k$\textit{\ there is a prime number }$p>k,$\textit{\ }$p\equiv 7%
\func{mod}8,$ \textit{such that all the non-negative integers up to }$k$%
\textit{\ are quadratic residues modulo }$p$\textit{. There are infinitely
many such primes }$p$\textit{\ for any given }$k$\textit{, they are
necessarily larger than }$2k$\textit{. \ \ \ \ }$\square $

\bigskip

From this we derive a key fact about the sets $C_{k}$, noting that these are
obtained from $\left\{ 1,2,...,k\right\} $ by symmetrization via the
adjunction of negatives and reciprocals:

\bigskip

\textbf{Symmetrized Chain Lemma}\textit{\ For every positive integer }$k$%
\textit{\ there is a prime number }$p>k,$\textit{\ }$p\equiv 7\func{mod}8,$ 
\textit{such that all the positive numbers in }$C_{k}\subseteq 
\mathbb{Z}
_{(p)}$ \textit{are quotients }$a/b$ \textit{of quadratic residues }$a,b$ 
\textit{modulo }$p$, \textit{i.e. all positive members of }$C_{k}$ \textit{%
are mapped to non-zero squares in }$\mathbf{F}_{p}$ \textit{by the canonical
surjection }$%
\mathbb{Z}
_{(p)}\longrightarrow \mathbf{F}_{p}.$

\textbf{Proof} \ By the Kustaanheimo Chain Theorem there is a prime number $%
p>k^{2},$\textit{\ }$p\equiv 7\func{mod}8,$ such that all the non-negative
integers up to $k^{2}$\ are quadratic residues modulo $p.$ $\ \ \ \ \ \ \ \
\ \ \ \ \ \ \ \ \ \ \ \ \ \ \ \ \ \ \ \ \ \ \ \ \ \ \ \ \ \ \ \ \ \ \ \ \ \
\ \ \ \ \ \ \ \ \ \ \ \ \ \ \ \ \ \ \ \ \ \ \ \ \ \ \square $

\bigskip

\bigskip

\textbf{2.4 Density and conclusion of proof of the theorem\bigskip }

By the \textit{coefficients} of a linear transformation of $%
\mathbb{R}
^{n}$ we mean the entries of its standard matrix representation.

In this Section we deal mainly with the metric space structure on the set of
all linear transformations of $%
\mathbb{R}
^{4}$ induced by the Euclidean norm of standard matrix representations as
briefly introduced in Section 1. In this metric space, the distance between
two transformations $T,Q$ is the norm of their difference, $\left\Vert
T-Q\right\Vert .$

In $\mathcal{L}_{+}^{\uparrow }=\mathcal{L}_{+}^{\uparrow }%
\mathbb{R}
$ the norm of any space rotation is $2$ and the norm of the basic boost $%
B_{\alpha }$ is $\alpha +\alpha ^{-1}$. If $T$ is any linear transformation
of $%
\mathbb{R}
^{4}$ and $R$ is any space rotation, then $\left\Vert T\right\Vert
=\left\Vert RT\right\Vert =\left\Vert TR\right\Vert .$ The norm of a proper
orthochronous Lorentz transformation $T=RSB_{\alpha }S$ in $\mathcal{L}%
_{+}^{\uparrow }$, where $R,S$ are space rotations, is $\alpha +\alpha
^{-1}\geq 2,$ and this is also the norm of the inverse transformation $%
T^{-1}.$

In any metric space, a set $A$ of elements\ is said to be $\epsilon $-%
\textit{dense} \textit{in} a set of elements $B$\ (where $\epsilon $ is any
positive real) if for every $b\in B$ there is an $a\in A\cap B$ at distance
less than $\epsilon $ from $b$. Note that this does not require $A$ to be a
subset of $B$, but $A$ is $\epsilon $-dense in $B$ if and only if $A\cap B$
is $\epsilon $-dense in $B.$ Note that if $A$ is $\epsilon $-dense in a set $%
B$, and also in a set $C$, then $A$ is $\epsilon $-dense in $B\cup C$, but
the converse implication generally fails. Density in the usual sense means $%
\epsilon $-density for all $\epsilon >0.$

\bigskip

\textbf{Approximation Lemma 1 \ }\textit{For every positive integer }$k$%
\textit{\ the set }$C_{k^{2}}$ \textit{is} $(1/k)$\textit{-dense in the real
interval} $[-k,k].$

\textbf{Proof \ }This follows from the inclusion

\[
\left\{ i/k:i\in 
\mathbb{Z}
\text{ \ }-k^{2}\leq i\leq k^{2}\right\} \subseteq C_{k^{2}}\cap \lbrack
-k,k] 
\]

$\square $

\bigskip

\textbf{Approximation Lemma 2 }\textit{Let }$\epsilon >0\mathit{\ }$\textit{%
and let K be the union of a finite number of non-trivial compact real
intervals. Then there is a positive integer l such that for all integers }$%
k\geq l$ the set $C_{k}$ \textit{is }$\epsilon $-\textit{dense in} $K$.

\textbf{Proof}\textit{\ \ }Approximation Lemma 1. \ \ \ $\square $

\bigskip

\textbf{Approximation Lemma 3}\textit{\ \ Let }$\epsilon >0$ \textit{and let 
}$\mathbf{u}$\textit{\ be any of the three standard space unit vectors. Then
there is a positive integer }$l$\textit{\ such that for all integers }$k\geq
l$ $\ $\textit{the set }$\mathcal{L}_{+}%
\mathbb{R}
C_{k}$ \textit{is }$\epsilon $-\textit{dense within the set of space
rotations around }$\mathbf{u}$.

\textbf{Proof \ }Suppose $\mathbf{u}=(0,1,0,0)$. In the other cases the
proof follows by conjugation.

The identity transformation and the half-turn around $\mathbf{u}$ have their
coefficients in $C_{k}$.

There is some $M>0$ such that for all basic rotations $R_{\alpha }$ with $%
M<\alpha $ the distance between $R_{\alpha }$\ and the identity is less than 
$\epsilon .$

There is some $0<\mu <M$ such that for all basic rotations $R_{\alpha }$
with $0<\alpha <\mu $ the distance between $R_{\alpha }$ the and the
half-turn is less than $\epsilon .$

Let $K=[-M,-\mu ]\cup \lbrack \mu ,M].$ By uniform continuity of the map $%
\alpha \mapsto R_{\alpha }$ on the compact set $K$, there is $\delta >0$
such that for all $\alpha ,\gamma \in K,$ $\left\vert \alpha -\gamma
\right\vert <\delta $ implies $\left\Vert R_{\alpha }-R_{\gamma }\right\Vert
<\epsilon $.

By Approximation Lemma 2, there is a positive integer $l$ such that for all
integers $k\geq l$ the set $C_{k}$ is $\delta $-dense in $K$.

Let $k\geq l$ and let $R_{\alpha }$ be a basic space rotation with $\alpha
\in K.$ By $\delta $-density there is a $\gamma \in C_{k}\cap K$ such that $%
\left\vert \alpha -\gamma \right\vert <\delta $. We have $\left\Vert
R_{\alpha }-R_{\gamma }\right\Vert <\epsilon .$ \ \ \ \ \ \ \ \ \ \ \ \ \ \
\ \ \ \ \ \ \ \ \ \ \ \ \ \ \ \ \ \ \ \ \ \ \ \ \ \ \ \ \ \ \ \ \ \ \ \ \ \
\ \ \ \ \ \ \ \ \ \ \ \ \ \ \ \ \ \ \ \ \ \ \ \ \ \ \ \ \ \ \ \ \ \ \ \ \ \
\ \ \ \ \ \ \ \ \ \ \ \ \ \ \ \ \ \ \ \ $\square $

\bigskip

\textbf{Approximation\ Lemma 4 }\textit{Let }$\epsilon >0$ \textit{and }$%
M>0. $ \textit{There is a positive integer }$l$\textit{\ such that for all
integers }$k\geq l$ $\ $\textit{the set }$\mathcal{L}_{+}%
\mathbb{R}
C_{k}$ \textit{is }$\epsilon $-\textit{dense within the set of basic boosts
of norm not exceeding }$M$\textit{.}

\textbf{Proof} \ There is some $0<\mu <M$ such that all basic boosts $%
B_{\alpha }$ with $\alpha <\mu $ have norm greater than $M.$\ 

Let $K=[\mu \cup M].$ By uniform continuity of the map $\alpha \mapsto
B_{\alpha }$ on the compact interval $K$, there is $\delta >0$ such that for
all $\alpha ,\gamma \in K,$ $\left\vert \alpha -\gamma \right\vert <\delta $
implies $\left\Vert B_{\alpha }-B_{\gamma }\right\Vert <\epsilon $.

By Approximation Lemma 2, there is a positive integer $l$ such that for all
integers $k\geq l$ the set $C_{k}$ is $\delta $-dense in $K$.

Let $k\geq l$ and let $B_{\alpha }$ be a basic boost with $\alpha \in K.$ By 
$\delta $-density there is a $\gamma \in C_{k}\cap K$ such that $\left\vert
\alpha -\gamma \right\vert <\delta $. We have $\left\Vert B_{\alpha
}-B_{\gamma }\right\Vert <\epsilon .$ \ \ \ \ \ \ \ \ \ \ \ \ \ \ \ \ \ \ \
\ \ \ \ \ \ \ \ \ \ \ \ \ \ \ \ \ \ \ \ \ \ \ \ \ \ \ \ \ \ \ \ \ \ \ \ \ \
\ \ \ \ \ \ \ \ \ \ \ \ \ \ \ \ \ \ \ \ \ \ \ \ \ \ \ \ \ \ \ \ \ \ \ \ \ \
\ \ \ \ \ \ \ \ \ \ \ \ \ \ \ $\square $

\bigskip

The above lemmas combine to yield the following

\bigskip

\textbf{Approximation Lemma 5 \ }\textit{Let }$\epsilon >0$ \textit{and }$%
M>0.$ \textit{There is a positive integer }$l$\textit{\ such that for all
integers }$k\geq l$ $\ $\textit{the set }$\mathcal{L}_{+}%
\mathbb{R}
C_{k}$ \textit{is }$\epsilon $-\textit{dense within each of the three
elementary space rotation groups around the standard space unit vectors, and
also within the set of basic boosts of norm not exceeding }$M$\textit{.} \ \
\ \ \ $\square $

\bigskip

\textbf{Approximation Lemma 6 \ }\textit{Let }$\epsilon >0$ \textit{and }$%
M>0.$ \textit{There is a positive integer }$l$\textit{\ such that for all
integers }$k\geq l$ $\ $\textit{the set }$\mathcal{L}_{+}%
\mathbb{R}
C_{k}$ \textit{is }$\epsilon $-\textit{dense within the set of orthochronous
proper Lorentz transformations of norm not exceeding }$M.$

\textbf{Proof}\textit{\ \ }By a compactness and uniform continuity argument,
there is a $\delta >0$ such that for all linear transformations%
\[
T_{1},...,T_{10},Q_{1},...,Q_{10} 
\]%
of $%
\mathbb{R}
^{4}$ of norm not exceeding $M,$ if $\left\Vert T_{i}-Q_{i}\right\Vert
<\delta $ for all $1\leq i\leq 10$, then we have for the distance of products%
\[
\left\Vert T_{1}...T_{10}-Q_{1}...Q_{10}\right\Vert <\epsilon 
\]

By Approximation Lemma 5, there is a positive integer $l$\ such that for all
integers $k\geq l$ $\ $the set $\mathcal{L}%
\mathbb{R}
C_{k}$ is $\delta $-dense within each of the three elementary space rotation
groups around the standard space unit vectors, and also within the group of
basic boosts of norm not exceeding\textit{\ }$M$\textit{.}

Let $T$ be any orthochronous proper Lorentz transformation, $T=RSB_{\alpha
}S^{-1}$ where $R,S$ are space rotations, and suppose that the norm $\alpha
+\alpha ^{-1}$ of $T$ is at most $M.$ We can factorize $R$ as a composition $%
R_{1}R_{2}R_{3}$ of three elementary space rotations, and factorize also as $%
S$ as the composition $R_{4}R_{5}R_{6}$ of three elementary space rotations.
For convenience write $R_{7}=R_{6}^{-1},R_{8}=R_{5}^{-1},R_{9}=R_{4}^{-1}$,
so that%
\[
T=R_{1}...R_{6}B_{\alpha }R_{7}R_{8}R_{9} 
\]%
By $\delta $-density, there are elementary space rotations $R_{1}^{\prime
},...,R_{9}^{\prime }$ with coefficients in $C_{k}$ and a basic boost $%
B^{\prime }$ also in $\mathcal{L}_{+}%
\mathbb{R}
C_{k}$ and of norm at most $M$, such that for each $0\leq i\leq 9$ the
elementary rotations $R_{i}$ and $R_{i}^{\prime }$ have a common fixed
standard space unit vector, and the distances%
\[
\left\Vert R_{1}^{\prime }-R_{1}\right\Vert ,...,\left\Vert R_{9}^{\prime
}-R_{9}\right\Vert ,\left\Vert B^{\prime }-B_{\alpha }\right\Vert 
\]%
are all less than $\delta .$ Then the distance of $T$ from the composition $%
R_{1}^{\prime }...R_{6}^{\prime }B_{\alpha }R_{7}^{\prime }R_{8}^{\prime
}R_{9}^{\prime }$ is at most $\epsilon $. \ \ \ \ \ \ \ \ \ $\square $

\bigskip

\textbf{Conclusion of proof of Theorem 1} Given\textit{\ }$\epsilon >0$ and%
\textit{\ }$M>0,$ by Approximation Lemma 6 we can take a positive integer%
\textit{\ }$k$\textit{\ }such that the set $\mathcal{L}_{+}%
\mathbb{R}
C_{k}$ is $\epsilon $-dense within the set of orthochronous proper Lorentz
transformations of norm at most $M.$

By the Symmetrized Chain Lemma, there is a prime $p,$ 
\[
p>32k^{16}>k,\text{ \ }p\equiv 7\func{mod}8 
\]%
such that the canonical surjection $%
\mathbb{Z}
_{(p)}$ $\longrightarrow \mathbf{F}_{p}$ maps all of $C_{k}$ inside the
group of non-zero squares of $\mathbf{F}_{p}.$ Then the canonical surjection 
$\mathcal{L}_{+}%
\mathbb{R}
\mathbb{Z}
_{(p)}\longrightarrow \mathcal{L}_{+}\mathbf{F}_{p}$ maps all of $\mathcal{L}%
_{+}^{\uparrow }%
\mathbb{R}
C_{k}$ inside $\mathcal{L}_{+}^{\uparrow }\mathbf{F}_{p}.$ Let $A=\mathcal{L}%
_{+}^{\uparrow }%
\mathbb{R}
C_{k}.$

Define the retraction $f:\mathcal{L}_{+}^{\uparrow }%
\mathbb{R}
\longrightarrow A$ by associating to each orthochronous proper Lorentz
transformation $T$ a member of$\ A$ at minimum distance from $T.$ Then apply
the Local Isomorphism Lemma of 2.3. \ \ \ $\square $

\bigskip

\bigskip

\bigskip

\bigskip \textbf{3}\bigskip\ \textbf{Extended Lorentz groups and line
reflections}

Given $\epsilon >0$ and $M>0$ the retraction of the proper orthochronous
Lorentz group $\mathcal{L}_{+}^{\uparrow }%
\mathbb{R}
$ provided by Theorem 1 can in fact be defined also on the other three
connected components of the extended Lorentz group, with the retract $A$
being locally isomorphic to a subset $Y$ of the \textit{extended Lorentz
group} $\mathcal{L}\mathbf{F}_{p}$ of all bijective linear transformations
of $\mathbf{F}_{p}^{4}$ preserving the Minkowski norm. The local isomorphism
between $A$ and $Y$ can also be seen to preserve several other properties of
transformations besides the property of being orthochronous.

\bigskip

As over $%
\mathbb{R}
$, over any field $\mathbf{F}$\ (where $-1$ is not a square) the proper
Lorentz group is an index $2$ subgroup of the extended group $\mathcal{L}%
\mathbf{F}$, and we denote by $\mathcal{L}_{-}\mathbf{F}$ the coset
consisting of the transformations of determinant $-1$, called \textit{%
improper} transformations. Among these we have in particular the \textit{%
time reversal} transformation $\tau $ given by 
\[
\tau (t,x,y,z)=(-t,x,y,z) 
\]%
Similarly to orthochronous transformations in $\mathcal{L}_{+}\mathbf{F,}$ a
transformation in $\mathcal{L}_{-}\mathbf{F}$ is also called o\textit{%
rthochronous }if it maps $(1,0,0,0)$ to a vector $(t,x,y,z)$ where $t$ is a
non-zero square in $\mathbf{F}$. The set of orthochronous improper Lorentz
transformations is denoted by $\mathcal{L}_{-}^{\uparrow }\mathbf{F}$.

\bigskip

Among all improper Lorentz transformations, a fundamental role is played by
line reflections\textit{\ }in time-like or space-like lines, i.e. $1$%
-dimensional subspaces of $\mathbf{F}^{4}$ consisting of the scalar
multiples of a vector $\mathbf{a}\in \mathbf{F}^{4}$ with Minkowski norm $%
\mu (\mathbf{a})$ equal to $1$ or $-1$. In the classical case $\mathbf{F=%
\mathbb{R}
}$ this role was recently emphasized by Urbantke in [U]. Over any field $%
\mathbf{F}$, \textit{reflection} in the line of $\mathbf{a=(}%
a_{0},a_{1},a_{2},a_{3})\in \mathbf{F}^{4}$ (where $\mu (\mathbf{a})=\pm 1$)
is the improper Lorentz transformation with matrix%
\[
2\mu (\mathbf{a})[\mathbf{a}^{\mathsf{\top }}\tau (\mathbf{a})-\mathbf{I]} 
\]%
where in the square bracket the matrices $\mathbf{a}^{\mathsf{\top }},\tau (%
\mathbf{a})$ and $\mathbf{I}$ are $4\times 1,1\times 4$ and $4\times 4,$
respectively, and $\tau $ is time reversal.

\bigskip

\bigskip

\textbf{Theorem 2 }\ \textit{For every} $\epsilon >0$ \textit{and }$M>0$ 
\textit{the extended Lorentz group} $\mathcal{L}%
\mathbb{R}
$ \textit{has a retraction}$\ f:\mathcal{L}%
\mathbb{R}
\longrightarrow A,$ $f^{2}=f,$ \textit{onto a finite subset} $A\subseteq 
\mathcal{L}%
\mathbb{R}
$ \textit{satisfying}

\[
\left\Vert T-f(T)\right\Vert <\epsilon 
\]%
\textit{for all Lorentz transformations} $T\in \mathcal{L}%
\mathbb{R}
$ \textit{of norm not exceeding} $M$, \textit{and such that} \textit{for
some prime number} $p\equiv 7\func{mod}8$ \textit{the retract} $A$\textit{\
is locally isomorphic to a subset of }$\mathcal{L}\mathbf{F}_{p}$\textit{\
via a local isomorphism }$\lambda $, \textit{where the retraction }$f$ 
\textit{and the local isomorphism }$\lambda $\textit{\ satisfy the following
conditions:}

(i) $f$ \textit{maps each of the }$4$ \textit{connected components of }$%
\mathcal{L}%
\mathbb{R}
$ \textit{to its intersection with }$A,$

(ii) $\lambda $\textit{\ maps each of the intersection sets }%
\[
A\cap \mathcal{L}_{+}^{\uparrow }%
\mathbb{R}
,\text{ \ }A\cap \mathcal{L}_{+}^{\downarrow }%
\mathbb{R}
,\text{ \ }A\cap \mathcal{L}_{-}^{\uparrow }%
\mathbb{R}
,\text{ \ }A\cap \mathcal{L}_{-}^{\downarrow }%
\mathbb{R}
\]%
\textit{\ respectively into the corresponding cosets}%
\[
\mathcal{L}_{+}^{\uparrow }\mathbf{F}_{p},\text{ \ }\mathcal{L}%
_{+}^{\downarrow }\mathbf{F}_{p},\text{ \ }\mathcal{L}_{-}^{\uparrow }%
\mathbf{F}_{p},\text{ \ }\mathcal{L}_{-}^{\downarrow }\mathbf{F}_{p} 
\]

(iii) $\lambda $ \textit{maps space rotations to space rotations, boosts to
boosts, line reflections to line reflections. \ \ \ \ \ \ }$\square $

\bigskip

\bigskip

The extended retraction and local isomorphism of Theorem 2 are constructed
similarly to those of Theorem 1. Generating the extended finite Lorentz
group $\mathcal{L}\mathbf{F}_{p}$ by reflections now leads to a
representation of the extended finite Lorentz group $\mathcal{L}\mathbf{F}%
_{p}$ as a quotient of the group of all Minkowski norm preserving
automorphisms of the module $%
\mathbb{Z}
_{(p)}^{4}.$

\bigskip

\bigskip

\bigskip

\textbf{4} \textbf{Remarks on finiteness and approximation}

\bigskip

It has been noted (see Beltrametti and Blasi [BB2]) that, as opposed to $%
\mathcal{L}_{+}^{\uparrow }\mathbf{%
\mathbb{R}
}$ and $\mathcal{L}^{\uparrow }\mathbf{%
\mathbb{R}
=}\mathcal{L}_{+}^{\uparrow }\mathbf{%
\mathbb{R}
}\cup \mathcal{L}_{-}^{\downarrow }\mathbf{%
\mathbb{R}
},$ the set of orthochronous Lorentz transformations over a finite field
does not constitute a group. For example, take any prime number $p\equiv 7%
\func{mod}8$ and such that $3$ is a quadratic residue modulo $p.$ The first
positive integer $q$ such that $q+1$ is not a quadratic residue modulo $p$
is even, and $q/2$ is then a quadratic residue. Let $\alpha $ and $\gamma $
be non-zero squares in $\mathbf{F}_{p}$ such that the elements of $\mathbf{F}%
_{p}$ corresponding to the integers $2$ and $q/2$ are $\alpha ^{2}$ and $%
\gamma ^{2}$ respectively. Then the basic boosts $B_{\alpha }$ and $%
B_{\gamma }$ are orthochronous but $B_{\alpha }B_{\gamma }=B_{\alpha \gamma
} $ is not. However, for all orthochronous transformations $T_{1},T_{2}$ in
the set $Y\subseteq \mathcal{L}_{+}\mathbf{F}_{p}$ of Theorem 1, the
construction of the local isomorphism between $A\subseteq \mathcal{L}%
_{+}^{\uparrow }\mathbf{%
\mathbb{R}
}$ and $Y$ implies that $T_{1}T_{2}$ is also orthochronous. Thus the
phenomenon of an antichronous product of two orthochronous parallel boosts
arises only when the factors are not confined to the set $Y$ which in the
Lorentz group over $\mathbf{F}_{p}$ represents, in the sense of local
isomorphism, the retract $A\subseteq \mathcal{L}_{+}^{\uparrow }\mathbf{%
\mathbb{R}
.}$

\bigskip

In contrast to parallel boost groups in $\mathcal{L}\mathbf{%
\mathbb{R}
,}$ in $\mathcal{L}\mathbf{F}_{p}$ the boost group%
\[
\{RB_{\alpha }R^{-1}:\alpha \neq 0,\text{ }\alpha \text{ square in }\mathbf{F%
}_{p}\} 
\]%
is cyclic for any space rotation $R$, and it is isomorphic to the
multiplicative subgroup of non-zero squares in $\mathbf{F}_{p}$ via the map $%
\alpha \mapsto RB_{\alpha }R^{-1}$. The finite model then involves in each
space direction the existence of a boost from which all parallel boosts in
that direction are obtainable by repeated application.

\bigskip

As in the real number based model, over a finite field $\mathbf{F}_{p}$ as
well the \textit{velocity} $v_{\alpha }$ associated with the basic boost $%
B_{\alpha }$ is $(\alpha -\alpha ^{-1})/(\alpha +\alpha ^{-1})$ and for the
velocities of boosts $B_{\alpha }$ and $B_{\gamma }$ and of their
composition $B_{\alpha \gamma }$ we have%
\[
v_{\alpha \gamma }=\frac{v_{\alpha }+v_{\gamma }}{1+v_{\alpha }v_{\gamma }} 
\]%
In finite field models as well as over $%
\mathbb{R}
,$ velocity never equals $1.$ For $p\equiv 7\func{mod}8,$ if $\alpha ^{2}+1$
is a square in $\mathbf{F}_{p}$ (and this is always the case if $B_{\alpha }$
is in the set $Y\subseteq \mathcal{L}_{+}^{\uparrow }\mathbf{F}_{p}$ locally
isomorphic to the retract $A\subseteq \mathcal{L}_{+}^{\uparrow }\mathbf{%
\mathbb{R}
}$ as in\ Theorem 1), then $1-v_{\alpha }$ is a square in $\mathbf{F}_{p}$
i.e. in the non-transitive order $<_{p}$ on $\mathbf{F}_{p}$ considered by
Kustaanheimo in [K1, K2] and given by%
\[
x<_{p}y\Leftrightarrow y-x\text{ is a non-zero square in }\mathbf{F}_{p} 
\]%
we have%
\[
-1<_{p}v_{\alpha }<_{p}1 
\]%
Velocities $v_{\alpha }$ not constrained by these inequalities in $\mathbf{F}%
_{p}$ may only arise from boosts in $\mathcal{L}\mathbf{F}_{p}$ that do not
correspond, via the local isomorphism of Theorem 1, to boosts in the retract 
$A\subseteq \mathcal{L}_{+}^{\uparrow }\mathbf{%
\mathbb{R}
}$ which approximate all boosts of the real model of norm not exceeding $M$
i.e. real boosts of speed not exceeding $(1-4/M^{2})^{1/2}.$ As the bound $M$
can be as large as needed, this means that superluminal velocity ($%
1<_{p}v_{\alpha }$ in $\mathbf{F}_{p}$) cannot arise from boosts in $%
\mathcal{L}\mathbf{F}_{p}$ that correspond to boosts in $\mathcal{L%
\mathbb{R}
}$ with speed effectively distinguishable from the speed of light.

\bigskip

\bigskip

\textbf{References}

\bigskip

[A1]\ Y. Ahmavaara, Relativistic quantum theory as a group problem, II.
World geometry and the elementary particles, Annales Ac. Sc. Fennicae A VI
Physica 95 (1962) 51pp

\bigskip

[A2] Y. Ahmavaara, The structure of space and the formalism of relativistic
quantum theory, I-IV, J. Math. Phys. 4 (1965) 87-93, 6 (1965) 220-227, 7
(1966) 197-201, 7 (1966) 201-204

\bigskip

[BB1] E.G. Beltrametti, A.A. Blasi, Dirac spinors, covariant currents and
the Lorentz group over a finite field, Nuovo Cimento LVA/2 (1968) 301-317

\bigskip

[BB2] E.G. Beltrametti, A. Blasi, Rotation and Lorentz groups in a finite
geometry, J. Math. Phys. 9 (1968) 1027--1035

\bigskip

[BGZG] A.A. Blasi, F. Gallone, A. Zecca, V. Gorini, A causality group in
finite space-time, Nuovo Cimento 10A/1 (1972) 19-36

\bigskip

[C] H.R. Coish, Elementary particles in a finite world geometry, Phys. Rev.
114 - 1 (1959) 383-388

\bigskip

[D] L.E. Dickson, Determination of the structure of all linear homogeneous
groups in a Galois field which are defined by a quadratic invariant,
American J. Math. 21 (1899) 193-256

\bigskip

[Jo] H. Joos, Group-theoretical models of local-field theories, J. Math.
Phys. 5 (1964) 155-164

\bigskip

[J\"{a}] G. J\"{a}rnefelt, Reflections on a finite approximation to
euclidean geometry. Physical and astronomical prospects.\ Annales Ac. Sc.
Fennicae A I. Math.-Phys. 96 (1951) 1-43

\bigskip

[JK] G. J\"{a}rnefelt, P. Kustaanheimo, An observation on finite geometries,
in Proc. Skandinaviske Matematikerkongress i Trondheim 1949, 166-182

\bigskip

[K1] P. Kustaanheimo, A note on a finite approximation of the euclidean
plane geometry, Comment. Phys.-Math. Soc. Sc. Fenn. XV. 19 (1950) 1-11

\bigskip

[K2] P. Kustaanheimo, On the fundamental prime of a finite world, Annales
Ac. Sc. Fennicae A I. Math.-Phys. 129 (1952) 1-7

\bigskip

[M] V. Moretti, The interplay of the polar decomposition theorem and the
Lorentz group, in Lecture Notes of Seminario Interdisciplinare di Matematica
5-153 (2006) 18 pages, also arXiv:math-ph/0211047v1 at www.arxiv.org

\bigskip

[N] Y. Nambu, Field theory of Galois fields, in Field Theory and Quantum
Statistics, eds. J.A. Batalin et.al., Institute of Physics Publishing 1987,
pp. 625-636

\bigskip

[S] I.S. Shapiro, Weak interactions in the theory of elementary particles
with finite space, Nuclear Phys. 21 (1960) 474-491

\bigskip

[U] H.K. Urbantke, Lorentz transformations from reflections: some
applications, Found. Phys. Lett. 16 (2003) 111-117

\bigskip

[Y] Q.A.M. Yahia, Leptonic decays in finite space, Nuovo Cimento XXIX/2
(1963) 441-450

\end{document}